# Signal detection using biphotons and potential application in axion-like particle search

**Le Phuong Hoang[a], Binh Xuan Cao[a*]**

[a]School of Mechanical Engineering, Hanoi University of Science and Technology,
Ha Noi 100000, Vietnam

**Abstract:** This paper presents a new optical system for detecting light signals associated with the change in incoming photon number. The system employs quantum correlation of photon pairs created via spontaneous parametric down-conversion (SPDC). The signal, if present, will perturb the flux of the incident photon stream. The perturbed photon stream is first projected through a birefringent crystal where SPDC occurs, converting a single high-energy photon into a pair of low-energy photons. The photons in each pair eventually arrive at separate detectors. By examining the biphoton correlation using the probability distribution of the photons at the detectors, which varies depending on the displacement of the main "pump" photon stream and the change in the number of photons, the small optical displacement of the photon stream and its variance can be determined. The change in incident photon number, in other words, the presence of light signal does not influence the average of the measured optical displacement values. Nevertheless, the change in optical displacement measurement variance when the number of incident photons has changed detects the light signal. This optical setup enables the detection of light signals with low noise and remarkably high precision and sensitivity using quantum correlation. The proposed technique has potential application for axion-like particle search in experimental high energy physics.

*Keywords:* Photon detector, Quantum correlation, Spontaneous parametric down-conversion, Primakoff effect, Axion-like particle.

1. **Introduction**

Recently, measurements using quantum correlation have been investigated and are expected to yield a high-precision and sensitive method of determining the perturbation in the number of incident photons due to the photon-to-axion conversion process in a magnetic-field-permeated region while remaining minimally affected by noise from the surrounding environment [1–4]. The quantum correlation is investigated by examining the probability distributions of biphotons or photon pairs that are generated when the laser beam passes through a nonlinear medium, such as a birefringent crystal. At this point, the average of the positions of the two correlated photons in each pair is evaluated at the detectors, rather than the positions of the individual photons. This approach appears to be considerably more precise and reliable than the methods involving investigation of uncorrelated photons. Biphoton correlation is employed in numerous areas of science and technology [5–15]. However, for certain practical applications in experimental high energy physics, this quantum theory seems to be exotic, although quantum correlation is widely known to improve the accuracy of optical metrology to the Heisenberg scale [16–18]. Therefore, using biphoton position correlation in high-precision optics in general, the detection of light signals in particular is expected to yield a wide variety of automatic optical quantum equipment that can achieve Heisenberg-scale accuracy; this method is thus substantially more precise than the previous optical methods of signal detection that only utilize uncorrelated photons.

The introduction of Peccei-Quinn (PQ) symmetry leading to the absence of CP violations with strong interaction [19–21] has revealed exotic pseudoscalar particles widely known as axions [22,23]. These particles are of the PQ symmetry-breaking energy scale $f_a$. The value of $f_a$ was set based on two theoretical models (KSVZ [24,25] and DFSZ [26,27]) and some constraints from astrophysics [28-31] and cosmology [32–36] within the range of $10^9$–$10^{12}$ GeV, which is equivalent to the range of the axion mass $m_a$, i.e., $10^{-6}$–$10^{-3}$ eV. Modern experiments to detect cosmic axions [37–40] and solar axions [41–44] are based on the coupling between an axion and two photons, which is analogous to the conversion of a photon into an axion in the presence of a magnetic field. This is known as the inverse Primakoff effect, and is indicated by the introduction of the axion-magnetic field interaction Lagrangian:

$$\mathcal{L}_a = \frac{g_\gamma \alpha}{4\pi f_a} a F^{\mu\nu} \tilde{F}_{\mu\nu} = \frac{g_{a\gamma\gamma}}{4} a F^{\mu\nu} \tilde{F}_{\mu\nu} , \qquad (1)$$

where $g_\gamma$ is the model-dependent coupling coefficient, which was approximated as 0.36 in the DFSZ model and -0.97 in the KSVZ model, $\alpha = \frac{1}{137}$ is the fine structure constant, $a = \sqrt{2\rho_a}/m_a$ is the axion field with axion density $\rho_a = 0.5 \; GeV/cm^3$ [27], and $f_a = f_\pi m_\pi \sqrt{m_u m_d}[m_a(m_u + m_d)]^{-1}$ is the axion decay constant of the energy scale, where $m_u$ and $m_d$ are the masses of light quarks $u$ and $d$, respectively, and $m_\pi$ and $f_\pi$ represent the mass of pion $\pi$ and the decay constant, respectively [45]. The coupling constant $g_{a\gamma\gamma} = \frac{g_\gamma \alpha}{\pi f_a}$ is thus in the range $10^{-15}$–$10^{-11}$ GeV$^{-1}$. Another axion-searching method uses the coupling between an axion and a photon, which is achieved via photon-regeneration experiments known as the "light shining through a wall" method [46–51]. A small fraction of photons in a laser beam converts into axions when the beam passes through the first region with a magnetic field, and the generated axions then convert into photons when the beam passes through the second magnetic field region. Accordingly, the signal of the generated photons can be captured. Because this technique requires two conversion processes, the obtained signal is very weak for detecting QCD axions. This technique is more suitable for detecting axion-like particles (ALPs), which have stronger coupling with photons in a similar process. Thus, the energy scale parameter $f_a$ and the mass $m_a$ of ALPs are significantly smaller than those of axions.

This research presents a new method for detecting light signals by capturing the change in probability distribution of the incident correlated photons with two detectors while shifting the reflecting mirror along the optical axis. The system measures the displacement of the reflecting mirror along the optical axis and its variance. This measurement is based on the quantum correlation of the biphoton positions by counting the number of correlated photons reaching the left and right sides of the detectors when the incident photon number is changing. It was revealed that while the average of the measured values of mirror displacement remains the same irrespective of the change in incoming photons, the change in optical displacement measurement variance determines the presence of light signals. Owing to being highly sensitive to the small light signal detection, this system is promising to detect ALPs which is produced from aforementioned inverse Primakoff process in high magnetic field region. The mechanism of potential ALPs detection is reported in the discussion section of the manuscript.

## 2. Methodology and Computation

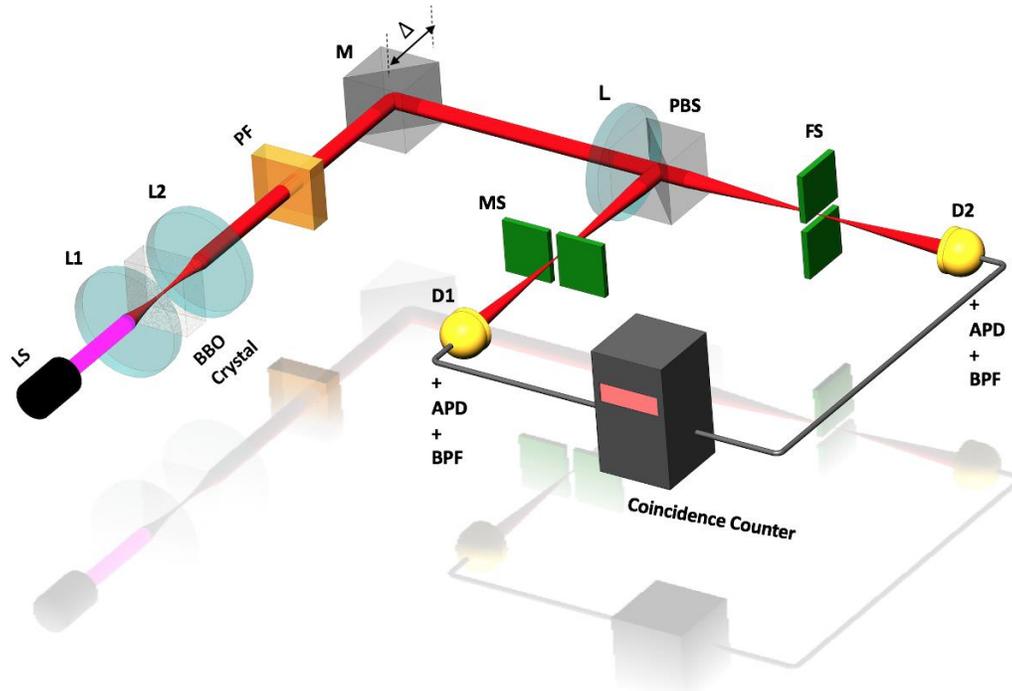

**Figure 1** Schematic of the optical system for light signal detection based on biphoton spatial correlation via displacement measurement of a movable mirror. The photon stream as a collimated beam from the laser source (LS) is focused by lens $L_1$ into a barium borate (BBO) crystal, where it experiences spontaneous parametric down-conversion, in which a photon is converted into two correlated photons with orthogonal polarizations and then collimated again by lens $L_2$. Next, the photon stream is reflected by a movable mirror (M) (moving direction is indicated) to the polarized beam splitter (PBS). The two correlated photons are separated by PBS, and are then detected by position-sensitive detectors D1 and D2, which are integrated in avalanche photo-diodes (APD) [4,5] at transverse positions $x_1$ and $x_2$, respectively. The pump filters (PF) and narrow band pass filters (BPF) in this setup are used to remove the uncorrelated as well as higher order correlated elements, and the lens (L) placed in front of the PBS is used to focus the fractional beam in the planes containing the fixed slit (FS) and the movable slit (MS), which are used to adjust the correlation distribution width [5,6]. In the experimental setup, the displacement of the reflecting mirror along the optical axis is exactly equal to the shift of the photon stream to PBS. The mirror is manually shifted along the optical axis. While the mirror is shifted, we ensure proper alignment such that the beam passes through the centers of the lens and slits. The detectors and PBS are always fixed during the movement of mirror.

The optical system is described in detail as follows. The collimated beam from the laser source (or pump beam) is focused by lens $L_1$ into a nonlinear medium (e.g., a Barium borate crystal), where incident single photons from the beam are converted into pairs of outgoing photons via spontaneous parametric down-conversion (SPDC) with perpendicular polarizations and then collimated again by lens $L_2$. The two photons are correlated due to conservation of energy and momentum, and express their correlation through the probability distribution of their positions at the detectors. Because we strictly obtain the second order generated correlated photons that have approximately twice the wavelength as the original uncorrelated photons owing to momentum and energy conservation, narrow band pass filters are used in the experiment. This also helps to maintain the consistency of measurement. Next, the correlated biphotons are split by a polarized beam splitter and directed toward two photon counter detectors with detection efficiency $\eta$ and two position-sensing detectors containing $\Omega$ pixels. The correlated position distribution varies depending on the displacement of the reflecting mirror $\Delta$. The correlated position distribution $p(x_1,x_2|\Delta)$ of the photon pair when the number of incoming photons is perturbed by small $\vartheta$ percent due to the presence of signal is given by [10–12],

$$p(x_1, x_2|\Delta) = \frac{1}{\pi\sigma\epsilon} \exp\left(\frac{-(x_1-x_2)^2}{2\sigma^2}\right) \times \exp\left(\frac{-(x_2+x_1-2\Delta)^2}{2\epsilon^2}\right) \times \eta(1-\vartheta) , \qquad (2)$$

where $\Delta$ is the displacement of the reflecting mirror along the optical axis, which will be measured; $\sigma$ is the waist of the pump beam, which is calculated from $\sigma = \sqrt{9L\lambda_o/10\pi}$, where $L$ is the thickness of the crystal, $\lambda_o$ is the wavelength of the pump beam, and $10\sigma/\Omega$ is the pixel width of the detector (detector width is $10\sigma$); $x_1$ and $x_2$ are the transverse positions of the photons in a given pair on detectors D1 and D2, respectively; and $\epsilon$ is the spatial correlation parameter or pump radius at the BBO crystal, which is focal spot with radius $\epsilon = 5\ \mu m$.

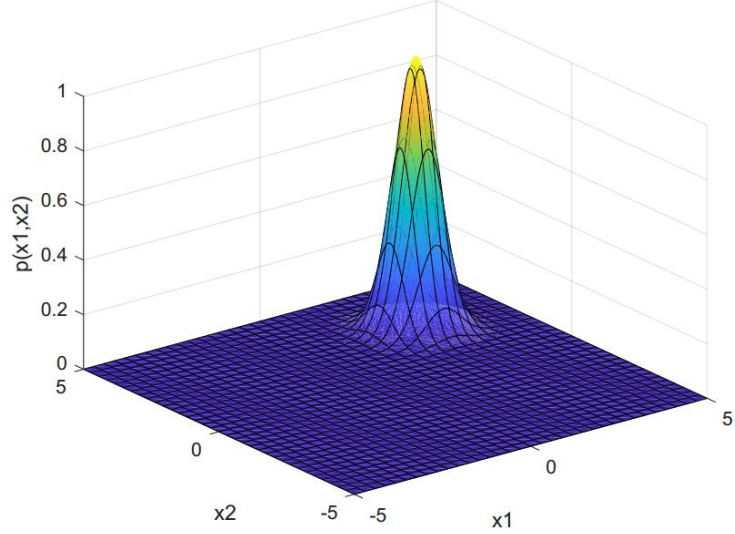

**Figure 2** Plot of distribution probability of correlated biphotons when the mirror is shifted by a displacement Δ along the optical axis, obtained by Eq. (2). According to the correlation, if a photon is captured at position $x_1$, the correlated photon has a peak possibility to arrive at position $x_2 = 2\Delta - x_1$. Overall, the peak probability of two correlated photons shifts to position $(x_1, x_2)$ such that $x_1 + x_2 = 2\Delta$.

In this method, the dependences of the correlated position distribution on the variances of $x_1 + x_2$ and $x_1 - x_2$ (which are defined as ϵ and σ, respectively, in Eq. (2) (σ is the beam waist at the slit plane, and ϵ is the beam radius at the crystal [6])) are examined to determine the displacement of the reflecting mirror along the optical axis. From the measured values of displacement and the variance of those values, the presence of ALPs that were generated from the photon-to-axion conversion can be determined. Based on Eq. (2), the following marginal distribution, which provides the probability distribution p(x|Δ) of the position of each photon in a biphoton pair on a detector, can be obtained [3]:

$$p(x|\Delta) = \int_{-\infty}^{+\infty} p(x, x_2|\Delta) dx_2 = \sqrt{\frac{2}{\pi(\epsilon^2+\sigma^2)}} \exp\left(\frac{-2(x-\Delta)^2}{\epsilon^2+\sigma^2}\right) \eta\ (1-\vartheta)\ . \qquad (3)$$

In the case of Δ = 0, the peak of the probability distribution of the position of each photon in a biphoton pair is located at $x = x_1 = x_2 = 0$. As the mirror is shifted by a displacement Δ ≠ 0, the peaks of the distributions shift toward $x_1$ and $x_2$ as long as $x_1 + x_2 = 2\Delta$, due to the correlation between the positions of the two photons in each pair. According to Eq. (2), the probabilities that a photon in a pair will reach the left ($p_-$) and right ($p_+$) sides of the corresponding detector are respectively given by [3],

$$p_- = \int_{-\infty}^0 p(x|\Delta)dx = \frac{1}{2}\left(1 - erf\left(\frac{\sqrt{2}\Delta}{\sqrt{\sigma^2+\epsilon^2}}\right)\right)\eta(1 - P_{\gamma\to a}) \approx \left(\frac{1}{2} - \sqrt{\frac{2}{\pi(\epsilon^2+\sigma^2)}}\Delta\right)\eta\ (1-\vartheta)$$

(4)

and

$$p_+ = \int_{-\infty}^0 p(x|\Delta)dx = \frac{1}{2}\left(1 + erf\left(\frac{\sqrt{2}\Delta}{\sqrt{\sigma^2+\epsilon^2}}\right)\right)\eta(1 - P_{\gamma\to a}) \approx \left(\frac{1}{2} + \sqrt{\frac{2}{\pi(\epsilon^2+\sigma^2)}}\Delta\right)\eta(1-\vartheta)\ .$$

(5)

We define $N_-$ and $N_+$ as the numbers of photons reaching the left and right sides of the detectors, respectively, where $N = N_- + N_+$ and $N$ is the total number of photons. By applying the maximum likelihood estimation to the above-mentioned probabilities, $\Delta$ can be obtained based on the data recorded by one detector [3]:

$$N_- \partial_\Delta lnp_- + N_+ \partial_\Delta lnp_+ = 0 \quad, \tag{6}$$

$$\Delta = \sqrt{\frac{\pi(\sigma^2+\epsilon^2)}{8}} \frac{N_+ - N_-}{N_+ + N_-} \quad. \tag{7}$$

We are interested in the average value of $\Delta$ calculated based on the probability distribution information obtained from the two detectors. For a large number of photons from a high-intensity laser beam, the average value of $\Delta$ can be measured by one detector. The uncertainty of this estimation is significant and can be computed using the Cramer–Rao bound of the Fisher information [52] about $\Delta$, $I(\Delta)$. $I(\Delta)$ indicates the amount of information that $x$ can reveal about $\Delta$ through its probability distribution. This quantity can be obtained as follows [3]:

$$I(\Delta) = N \sum_k \frac{(\partial_\Delta P_k)^2}{P_k} \quad, \tag{8}$$

where $P_k$ includes the probabilities of two photons arriving at the left side $P(-2|\Delta)$ and right side $P(2|\Delta)$ and the probability of one photon arriving at each side $P(0|\Delta)$ of a single detector [3]:

$$P(-2|\Delta) = \int_{-\infty}^{0} p_- dx \approx \left(\frac{1}{4} + \frac{1}{2\pi}arctan\left(\frac{\epsilon}{2\sigma} - \frac{\sigma}{2\epsilon}\right) - \Delta\sqrt{\frac{2}{\pi(\sigma^2+\epsilon^2)}}\right)\eta(1-\vartheta), \tag{9}$$

$$P(2|\Delta) = \int_{0}^{+\infty} p_+ dx \approx \left(\frac{1}{4} + \frac{1}{2\pi}\arctan\left(\frac{\epsilon}{2\sigma} - \frac{\sigma}{2\epsilon}\right) + \Delta\sqrt{\frac{2}{\pi(\sigma^2+\epsilon^2)}}\right)\eta(1-\vartheta) \tag{10}$$

and

$$P(0|\Delta) = \int_{-\infty}^{0} p_+ dx + \int_{0}^{+\infty} p_- dx \approx \frac{1}{2} - \frac{1}{\pi}\arctan\left(\frac{\epsilon}{2\sigma} - \frac{\sigma}{2\epsilon}\right)\eta(1-\vartheta) \quad. \tag{11}$$

$I(\Delta)$ can then be obtained as

$$I(\Delta) \approx \frac{16N}{(\epsilon^2+\sigma^2)(\pi+2\arcsin\xi)}\eta(1-\vartheta) \quad, \tag{12}$$

where $\xi = \frac{\epsilon^2-\sigma^2}{\epsilon^2+\sigma^2}$ is the correlation coefficient between two correlated photons. By assigning $\xi = 0$, the Fisher information for independent photons (or uncorrelated photons) $I_o(\Delta)$ can be obtained as follows [3]:

$$I_o(\Delta) \approx \frac{16N}{(\epsilon^2+\sigma^2)\pi}\eta(1-\vartheta) \quad. \tag{13}$$

The increase (or decrease) in the amount of information that the transverse position variable can provide about $\Delta$ when biphoton correlation is considered, compared with those when only pairs of independent photons are considered, is given by [3],

$$\frac{I(\Delta)}{I_o(\Delta)} = \frac{\pi}{\pi+2\arcsin\xi} \quad. \tag{14}$$

According to Eq. (14), more information about $\Delta$ with biphoton correlation is provided when $\xi < 0$, implying that $\epsilon < \sigma$. Thus, the photon stream as laser beam was focused tightly at BBO crystal, a lens with larger focal length was placed in front of the polarized beam splitter, and the pump filter was selected so that $\sigma$ was larger than $\epsilon$. Satisfying the Cramer–Rao bound [52] yields the following relation between I($\Delta$) and the minimum variance of $\Delta$, Var($\Delta$) [3]:

$$\text{Var}(\Delta) = \frac{1}{I(\Delta)} = \frac{(\epsilon^2+\sigma^2)(\pi+2\arcsin\xi)}{16N}\frac{1}{\eta(1-\vartheta)} . \tag{15}$$

Var($\Delta$) indicates the accuracy of $\Delta$ measurement; additionally, it represents the change percentage of incoming photon and thus indicates presence of light signals. Apparently, when the total number of photons arriving at the detectors increases, the measurement uncertainty decreases. Furthermore, Eq. (14) shows that employing biphoton correlation is more effective and more precise than using uncorrelated photons; this is because if a certain accuracy is achieved with N correlated incoming photons, $N \times \frac{\pi}{\pi+2\arcsin\xi}$ ($> N$) uncorrelated photons must be investigated to obtain the same accuracy [3]. The experiment is repeated several times to obtain numerous measured displacement values, with a fixed number of incoming photons. During each measurement, the displacement value $\Delta_i$ is obtained using Eq. (7) by employing the same method given in Ref. [3]; $\hat{\Delta}$ is the average value of obtained displacement after $n$ measurements. The variance of the measured displacement $\Delta$ over a number of measurements $n$ is computed by

$$\text{Var}(\Delta) = \frac{1}{n}\sum_{i=1}^{n}(\Delta_i - \hat{\Delta})^2 . \tag{16}$$

By setting this value equal to the minimum variance value obtained by the Cramer–Rao bound (Eq. 15), the presence of small light signals ($\vartheta \ll 1$) is indicated by the change in variance limit computation:

$$\delta Var(\Delta) = [Var(\Delta)](with\ signal) - [Var(\Delta)](without\ signal) = \frac{(\epsilon^2+\sigma^2)(\pi+2\arcsin\xi)}{16N\eta}\vartheta . \tag{17}$$

The detection signal-to-noise ratio, which considers the measurement resolution is given by:

$$\frac{S}{N} = \frac{\Delta}{\sqrt{Var(\Delta)}} = \Delta\sqrt{I(\Delta)} = \sqrt{\frac{2\pi}{N}}\sqrt{\frac{\eta(1-\vartheta)}{\pi+2arcsin\xi}}(N_+ - N_-) . \tag{18}$$

We calculated the signal-to-noise ratio by increasing the number of incoming photons based on Eqs. (18). The number of incoming photons increases as the laser power increases. The system settings are a maximum laser power of 5 W ($\lambda_o = 400$ nm), tunable laser pulse repetition rate of 400 kHz, detection efficiency $\eta$ of 85%, beam radius of 1 mm ($2w = 2$ mm), which can be changed by changing the aperture, and the focal spot at BBO crystal is 5 $\mu m$ which is also tunable using different lens. The mirror is manually shifted by $\Delta = 2.5$ mm. In addition, the number of detector pixels was $\Omega = 50$ and the thickness of the BBO crystal $L$ was 2 mm. We performed 1500 random displacement computations, meaning 1500 repetitions, at a specific number of incoming photons based on Eq. (7). With this setting, we acquire the total incoming photon flux dependence of signal-to-noise ratio shown in Fig. 3. Figure shows that the signal-to-noise ratio ranges from $6 \times 10^5$ to $2.9 \times 10^6$ when total flux of incident photons runs from $10^9$ to $7 \times 10^9$ photons per second. In actual experiments, the optical elements must be meticulously aligned to minimize errors induced by beam waists and deviations.

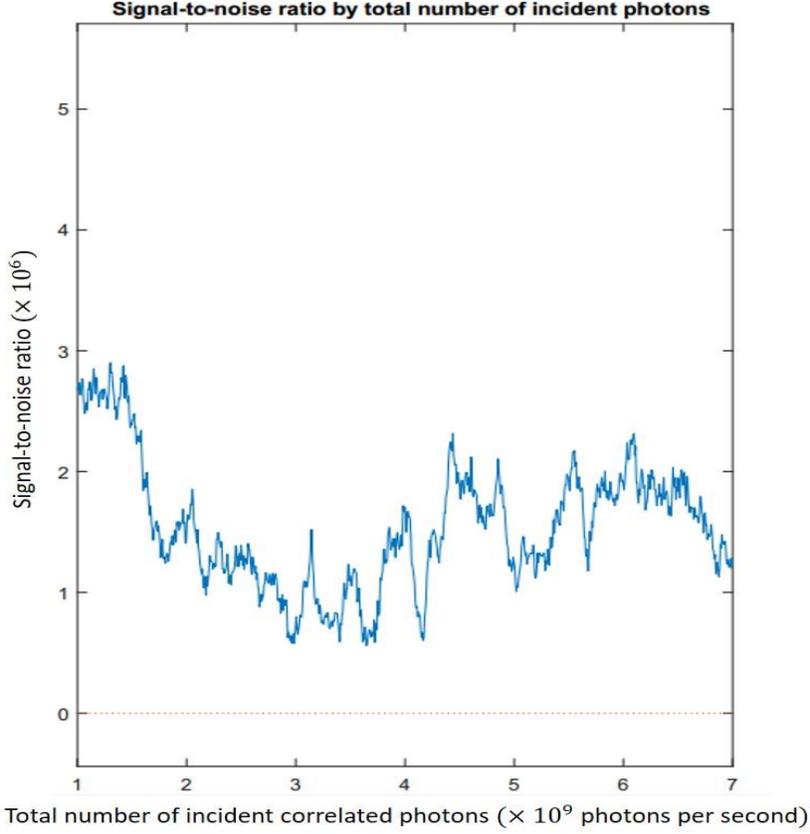

**Figure 3** Signal-to-noise ratio by total number of incident photons obtained using Eq. (19).

3. Discussion

Owning to being sensitive to the small change of signal with the signal-to-noise ratio in the range from $6 \times 10^5$ to $2.9 \times 10^6$ with respect to the flux of incident photons from $10^9$ to $7 \times 10^9$ photons per second (Fig. 3), we propose the potential application of the method on ALPs detection through inverse Primakoff process. To perform this detection experiment, one can set the optical system as shown in Fig. 4. The photons pass through a region that is permeated by a magnetic field perpendicular to the optical axis (Z-axis) after experiencing the SPDC process. The magnetic field region is built using a magnet system that contains two identical racetrack magnets (whose length $\zeta$ is described in [50]) to provide a large homogeneous magnetic field (Fig. 4). The magnetic lines of magnet 1 are perpendicular to those of magnet 2, and all magnetic lines are perpendicular to the beam propagation direction. The horizontal magnetic field induces the photon-to-axion conversion of horizontally polarized photons, and the vertical magnetic field induces the photon-to-axion conversion of vertically polarized photons. The change percentage in light signal is now equal to photon-to-axion conversion probability of the polarized photons in a parallel magnetic field is as follows [43]:

$$\vartheta = P_{\gamma \to a} = \frac{g_{a\gamma\gamma}^2}{4} \frac{\omega_\gamma}{\sqrt{\omega_\gamma^2 - m_a^2}} \left| \int B(z,t) e^{iqz} dz \right|^2 \approx \frac{1}{4} \left( g_{a\gamma\gamma} B \zeta \right)^2 , \qquad (19)$$

where $\omega_\gamma$ and $B$ represent the laser photon frequency and magnetic field amplitude, respectively, and $\zeta$ denotes the length of the region that is permeated by magnetic field $\boldsymbol{B}$. This approximation was made with the assumption that the magnetic field is homogeneous in the region and the ALP mass is negligibly small compared with the photon frequency ($m_a \ll \omega_\gamma$), since $m_a \sim 10^{-4}\ eV$, $\omega_\gamma \sim 1\ eV$, and $q = \frac{m_a^2}{2\omega_\gamma}$. Accordingly, the change in magnetic field amplitude $B$ is associated with the change in displacement measurement variance and thus demonstrate the efficiency of photon-to-axion conversion process. Similar to (17), we have:

$$\delta Var(\Delta) = [Var(\Delta)](B) - [Var(\Delta)](0) = \frac{(\epsilon^2+\sigma^2)(\pi+2\arcsin\xi)}{64N\eta}\left(g_{a\gamma\gamma}B\zeta\right)^2 . \tag{20}$$

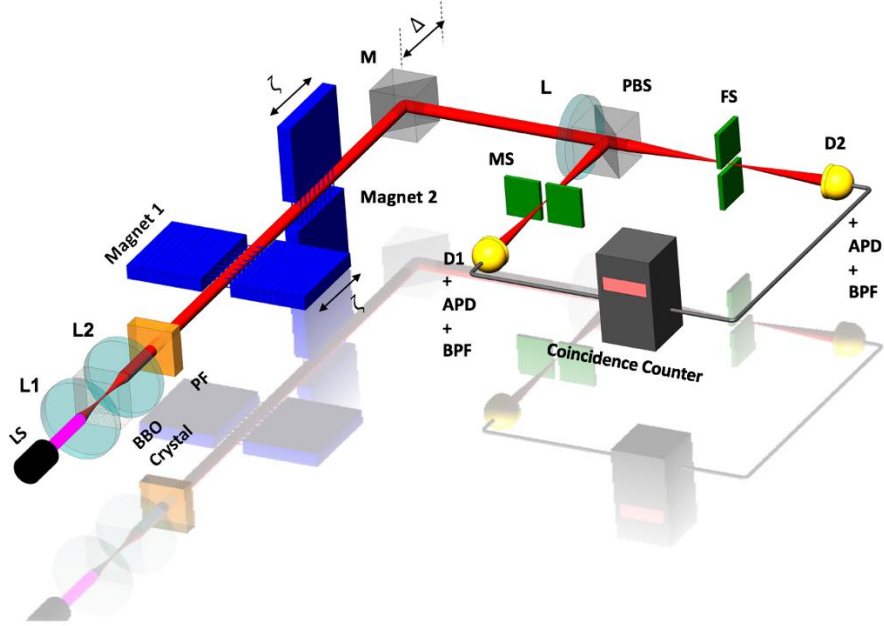

**Figure 4** Schematic of the enhanced optical system for ALPs detection based on biphoton spatial correlation via displacement measurement of a movable mirror.

Table I lists the potential systematic errors of the optical elements (laser source, photon detectors, and magnets) based on the specifications, as well as the alignment in an actual optical setup. The largest error ($\pm 0.4\%$) arises from the beam waists of each photon beam due to the inherent difficulties in placing the fixed slit at the focal position. In addition, the homogeneity of the magnetic field is maintained over a long distance using two identical racetrack magnets (two perpendicular magnetic field vectors) that can simultaneously convert both horizontally and vertically polarized photon beams induced by the SPDC process [50]. Furthermore, the optical paths of two correlated photon beams should be carefully considered to minimize errors in the flux of incident photons and thus reduce the noise. One factor that contributes to the uncertainty of $g_{a\gamma\gamma}$ is the photon-to-graviton conversion in the magnetic field permeated region based on linearized general relativity with the following conversion probability [53]:

$$P_{\gamma \to h} = 16\pi G B^2 \zeta^2 \quad , \tag{21}$$

where $G$ is the gravitational constant. The magnetic field converts both horizontally and vertically polarized photons irrespective of whether their direction is horizontal or vertical, as long as they are perpendicular to the beam propagation direction. Because the photon-to-graviton conversion probability is significantly smaller than the photon-to-axion conversion probability in current experimental conditions (i.e., B = 10 T, $\zeta$ = 9 m ($\mathcal{O}(10^{-33})$ compared with $(10^{-23})$), this graviton production probability is negligible. Equation (7) indicates that the measured displacement of the mirror does not depend on the photon-to-axion conversion rate, and both these values yield the same average displacement value irrespective of magnetic field presence. This result indicates that although the conversion decreases the number of incident photons, the ratio $\frac{N_+ - N_-}{N_+ + N_-}$ for each detector remains unchanged. We plan to conduct experiments with different reflecting mirror displacements and different magnetic field magnitudes, as well as with several types of photon counters that have better detection efficiency, to increase and emphasize the accuracy of the proposed model [54]. Furthermore, the system generating the magnetic field should be improved to maintain a strong magnetic field over a large area to realize the simulation results by optimizing the total length of the multi-magnet arrays and studying the materials of magnetic field sources (such as the wire material of single solenoids) as alternative candidates for single magnets [17,18]. Furthermore, we plan to enhance the laser system to perform stable measurements at higher powers to assess the efficiency of the proposed technique. The current experimental conditions do not allow for measurements at higher powers, because stably maintaining a high laser power is extremely difficult. The proposed technique should be considered a good candidate for the plethora of high-precision quantum devices that will be developed in the future to detect exotic particles.

**TABLE I Systematic errors in optical apparatuses**

| Source of error | Error (%) |
| --- | --- |
| Detection efficiency | ±0.3 |
| Beam waist | ±0.4 |
| Deviation of beam due to misalignment | ±0.1 |
| Magnetic field | ±0.1 |
| Repetition rate | ±0.1 |

### 4. Conclusion

We proposed a new technique for detecting light signals by treating the laser beam as a stream of photons undergoing biphoton quantum correlation. By measuring the number of correlated photons reaching the left and right sides of the detectors mapping the position distributions of the incident photons, the displacement of the reflecting mirror along the optical axis and its variance could be computed. The variance of displacement measurements indicates the presence of signals. The proposed technique can potentially lead to a plethora of axion and ALP detection systems using correlated biphotons that can probe the extreme range of these exotic particles.